\documentclass[11pt]{article}
\usepackage[a4paper,margin=2.4cm]{geometry}
\usepackage{amsmath,amssymb}
\usepackage{graphicx}
\usepackage{booktabs}
\usepackage{array}
\usepackage{placeins}
\usepackage[hidelinks]{hyperref}

\allowdisplaybreaks
\title{Vitrify Exact Learning: Gems, Issue 1 (November 2026)}
\author{Benedict W. J. Irwin\\[2pt] \normalsize Vitrify Labs}
\date{27 November 2026}
\newcommand{\el}{Vitrify Exact Learning}
\begin{document}
\maketitle

\begin{abstract}
Vitrify Exact Learning takes numerical values of a function and returns the linear differential equation with polynomial coefficients that annihilates it, with exact coefficients, together with the closed form that matches the values. This issue collects results obtained with it, and with integer relation searches (PSLQ) where stated, during November 2026, for which a search of the literature did not find the result. They are: a fourth-order differential equation, a partial differential equation and Appell, Kamp\'e de F\'eriet and Horn forms for the density of the distance between two random points in an ellipse; the PBE exchange energy of the hydrogen atom in the trilogarithm; the short-range (erf-attenuated) LDA exchange energy of the hydrogen atom as a $ {}_5F_5$, and of every density $n_0e^{-ar^p}$ with $3/p$ a whole number as a $ {}_{j+2}F_{j+2}$; the second column of the Broadhurst--Roberts determinant of six-Bessel moments at level 24 as Eichler integrals of a weakly holomorphic modular form; exact states of a harmonic trap with soft Coulomb and Lorentzian terms; and the period of the planar $L_p$ dual Minkowski problem as a Fox $H$ and Meijer $G$ function. A second part reports known results that were recovered from numbers alone, and the errors in published sources that the comparison found: a misprinted coefficient in a four-loop banana equation, one wrong parameter in each of four closed forms for three-dimensional lattice Green functions, and a series expansion of the ellipse density that does not converge to it. Every identity is a numerical check, stated with the number of digits to which it holds; none is proved here.
\end{abstract}

\tableofcontents

\FloatBarrier
\section{Introduction}

\el{}~\cite{el} is an algorithm that takes numerical values of a function on an interval and proceeds in three steps. It assumes that the function, in its own variable or after a change of variable, is annihilated by a linear differential operator with polynomial coefficients; it finds that operator from integrals of the data against test functions, makes its coefficients exact and tests it on data the fit did not read; and it solves the exact operator symbolically, keeping the closed form that matches the data. The method paper~\cite{el} describes the algorithm and its benchmarks: on noise-free data it recovers 45 of 53 classic targets, 108 of the 120 AI Feynman equations, 74.1\% of the Feynman problems of SRBench and 50 of 132 special-function targets. Where a result below names a constant rather than a function, it was found by integer relation searches with PSLQ~\cite{pslq}, run on the same numbers; we write \textquotedblleft Exact Learning + PSLQ\textquotedblright{} there.

This issue reports what the algorithm found when it was pointed at open or unfinished problems. Section~\ref{sec:new} gives six results that we did not find in the literature. Section~\ref{sec:known} gives known results that it recovered from numbers alone, because the comparison with the published forms exposed errors in them. Tables~\ref{tab:summary} and~\ref{tab:errata} summarise both.

\paragraph{Standard of claim.} Each identity below was checked numerically against a reference that does not use it, usually the defining integral computed directly in arbitrary precision; the precision is stated with each result. These are numerical identities, not proofs. Where we say a result was \emph{not found in the sources searched}, we mean that the searches described in each section, and in Section~\ref{sec:search}, did not find it; we do not claim it is new.

\paragraph{How the literature was searched.} For each result we read the closest papers in full; searched arXiv, OpenAlex, Crossref, zbMATH Open and general web search with several phrasings; and screened by title and abstract every work that OpenAlex lists as citing the closest reference. Table~\ref{tab:summary} lists the counts. We also asked a web search assistant for each closed form. Its answers were wrong in a recurring way, and we record those errors in Section~\ref{sec:search} because they are what a reader who searches for these results is likely to find.

\FloatBarrier
\section{Results not found in the sources searched}
\label{sec:new}

\FloatBarrier
\subsection{The distance between two random points in an ellipse}

Let $f(r,b)$ be the density of the distance between two independent uniform points in the ellipse $x^2 + y^2/b^2 < 1$, $0 < b < 1$. Stretching the ellipse to the unit disk writes it as an average over directions,
\begin{equation}
f(r,b) = \frac{2}{\pi}\int_0^{\pi/2}\frac{1}{s}\,f_{\rm disk}\!\left(\frac{r}{s}\right)d\theta,\qquad s = \sqrt{\cos^2\theta + b^2\sin^2\theta},
\label{eq:ell-avg}
\end{equation}
with $f_{\rm disk}(p) = \frac{4p}{\pi}\arccos\frac p2 - \frac{p^2}{\pi}\sqrt{4-p^2}$ for $p < 2$. The density has two pieces, $0 < r < 2b$ and $2b < r < 2$.

\paragraph{Result.} Given tables of $f$ at $b = \frac12, \frac13, \frac23, \frac34$, \el{} returned a fourth-order equation at each, with coefficients polynomial in $b$. Read across the four, they give one equation that holds on both pieces for every $b$:
\begin{multline}
r^4(r^2-4)(r^2-4b^2)\,f'''' + 2r^3(r^4-16b^2)\,f''' - r^2(3r^4-64b^2)\,f'' \\ + r(3r^4-128b^2)\,f' + (128b^2-3r^4)\,f = 0 .
\label{eq:ell-ode}
\end{multline}
At $b = 1$ it annihilates $f_{\rm disk}$. Given $f$ as a function of both variables, the algorithm returned the partial differential equation $r(b^2-1)f_{rb} + b(b^2-1)f_{bb} + rbf_r + (3b^2-1)f_b + bf = 0$. On the outer piece, with $\nu = 1 - \frac{r^2}{4}$ and $Y = \frac{b^2(4-r^2)}{r^2(1-b^2)}$, the density is one Appell function,
\begin{equation}
f = \frac{2\nu^2}{\pi\sqrt{1-b^2}}\,F_3\!\left(\tfrac12, \tfrac12, \tfrac32, \tfrac12; 3; \nu, Y\right),\qquad 2b < r < 2,
\end{equation}
and on the inner piece it is a Kamp\'e de F\'eriet function of $r^2/4$ and $1-b^2$,
\begin{equation}
f = \frac{2r}{b} - \frac{4r^2}{\pi}\sum_{j\ge0}\frac{(-\frac12)_j(\frac12)_j}{(\frac32)_j\,j!}\left(\frac{r^2}{4}\right)^{j} {}_2F_1\!\left(j+\tfrac32,\tfrac12; 1; 1-b^2\right),\qquad 0 < r < 2b,
\end{equation}
which near $r = 2b$ is better summed as a Horn $H_2$ series in $\nu$ and $\kappa = \frac{r^2(1-b^2)}{b^2(4-r^2)}$.

\begin{figure}[!htb]
\centering
\includegraphics[width=0.75\linewidth]{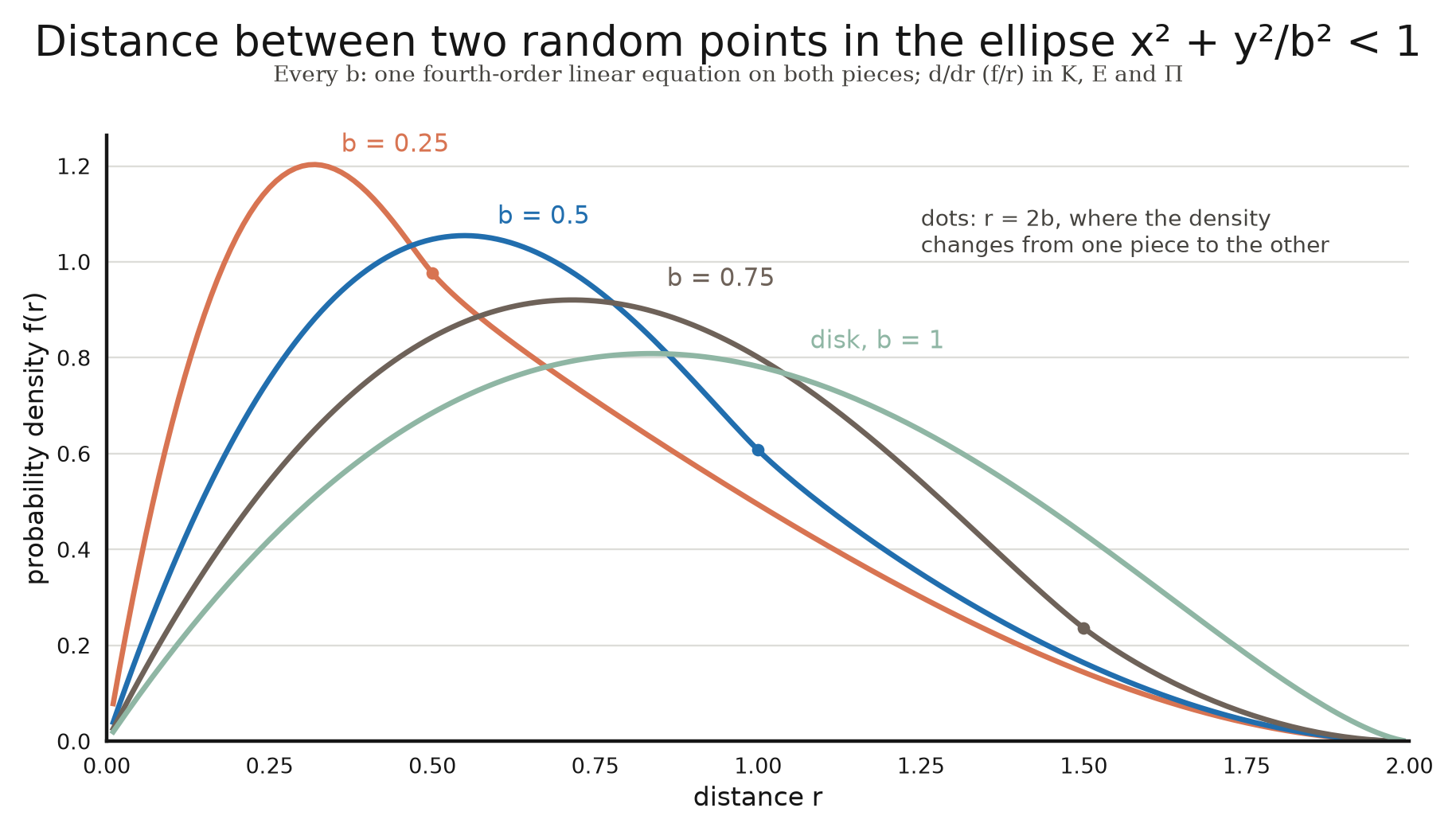}
\caption{The density $f(r,b)$ for $b = 1/4, 1/2, 3/4$ and the disk. Dots mark $r = 2b$, where the density changes from one piece to the other.}
\end{figure}

\paragraph{How it was found.} The equation~\eqref{eq:ell-ode} and the PDE came directly from the algorithm. The slope $S = \frac{d}{dr}(f/r)$ splits into two angular integrals; given each as a function of two variables, the algorithm returned them as an Appell $F_2$ (inner piece) and an Appell $F_1$ (outer piece). Integrating the $F_2$ once in $r$ from $f/r = 2/b$ at $r = 0$ gives the Kamp\'e de F\'eriet series, which satisfies the system of two partial differential equations the algorithm returned for its double sum. The $F_3$ and $H_2$ forms were then written by hand and checked.

\paragraph{Checks.} At 30 digits, against~\eqref{eq:ell-avg}: equation~\eqref{eq:ell-ode} leaves a relative residual below $2\times10^{-31}$ at $b = 3/10$, a value no run used; the $F_3$ and Kamp\'e de F\'eriet forms match 60-digit tables at four values of $b$ to $4.9\times10^{-60}$, and~\eqref{eq:ell-avg} at $b = 3/10$ and $7/10$ to $2.1\times10^{-30}$; the $H_2$ series meets the $F_3$ at $r = 2b$ in value and slope to $2\times10^{-31}$.

\paragraph{Literature.} Kellerer~\cite{kellerer} gives $f$ as a one-dimensional integral (his (A.10)) and the moments, and writes that the integral \textquotedblleft can not be solved in closed form, but can be expressed in terms of standard Legendre functions\textquotedblright, without giving that expression. Parry and Fischbach~\cite{pf} expand $f$ in the eccentricity (Section~\ref{sec:pf}). Ge\v{c}iauskas~\cite{gec}, Sheng~\cite{sheng}, Burgstaller and Pillichshammer~\cite{bp}, Lellouche and Souris~\cite{ls} and Garc\'ia-Pelayo~\cite{gp} (the spheroid) give general formulas, moments or bounds. We did not find the equation, the PDE or the closed forms in these, in the 28 and 77 works citing Parry--Fischbach and Kellerer, or elsewhere. The slope is closer to known results than the density: for a convex body $\frac{d}{dr}(f/r)$ is $-2L/A^2$ times the tail of the chord-length distribution under isotropic uniform random lines, with $L$ the perimeter and $A$ the area, and for the ellipse Kellerer gives that distribution as the integral (A.8). Written in complete elliptic integrals, on $2b < r < 2$ with $w = \frac{4-r^2}{4(1-b^2)}$,
\begin{equation}
S = -\frac{4(1-b^2)E(w) + 4b^2K(w) + \bigl(4b^2-(1+b^2)r^2\bigr)\Pi\!\left(1-\frac{r^2}{4}, w\right)}{\pi^2b^2\sqrt{1-b^2}},
\end{equation}
with $K$, $E$, $\Pi$ in the parameter convention, and a similar form on $r < 2b$. It matches the chord-length tail computed directly over random lines to 30 digits on both pieces; we record it as a convenience.

\FloatBarrier
\subsection{The PBE exchange energy of the hydrogen atom}

For the spin-polarised hydrogen density $n = e^{-2r}/\pi$ the PBE exchange energy~\cite{pbe} reduces, with $t = 4r/3$ and $c = \mu/(\kappa(6\pi)^{2/3})$, to $E_x^{\rm PBE} = E_x^{\rm LDA}(1 + \kappa - 4\kappa G(c))$, where $G(c) = \int_0^\infty t^2e^{-2t}/(1 + ce^t)\,dt$ and $E_x^{\rm LDA} = -\frac{81}{256}(6/\pi^2)^{1/3}$.

\paragraph{Result.} Given values of $G$ on $0.05 < c < 3$, \el{} returned the operator $c^3(c+1)D^4 + c^3D^3 - c(2c-1)D^2 + (6c-1)D - 8$ and $G = \frac{2}{27c}\, {}_4F_3(1,3,3,3;4,4,4;-1/c)$, which sums to $\frac14 - 2c - 2c^2\,\mathrm{Li}_3(-1/c)$. Hence, for every $\kappa$ and $\mu$,
\begin{equation}
E_x^{\rm PBE}(\mathrm{H}) = E_x^{\rm LDA}\left[1 + 8\kappa c\left(1 + c\,\mathrm{Li}_3(-1/c)\right)\right],
\end{equation}
which at PBE's $\kappa = 0.804$, $\mu = 0.2195149727645171$ is $-0.30594056823331236959$~hartree, against the exact $-5/16$.

\begin{figure}[!htb]
\centering
\includegraphics[width=0.75\linewidth]{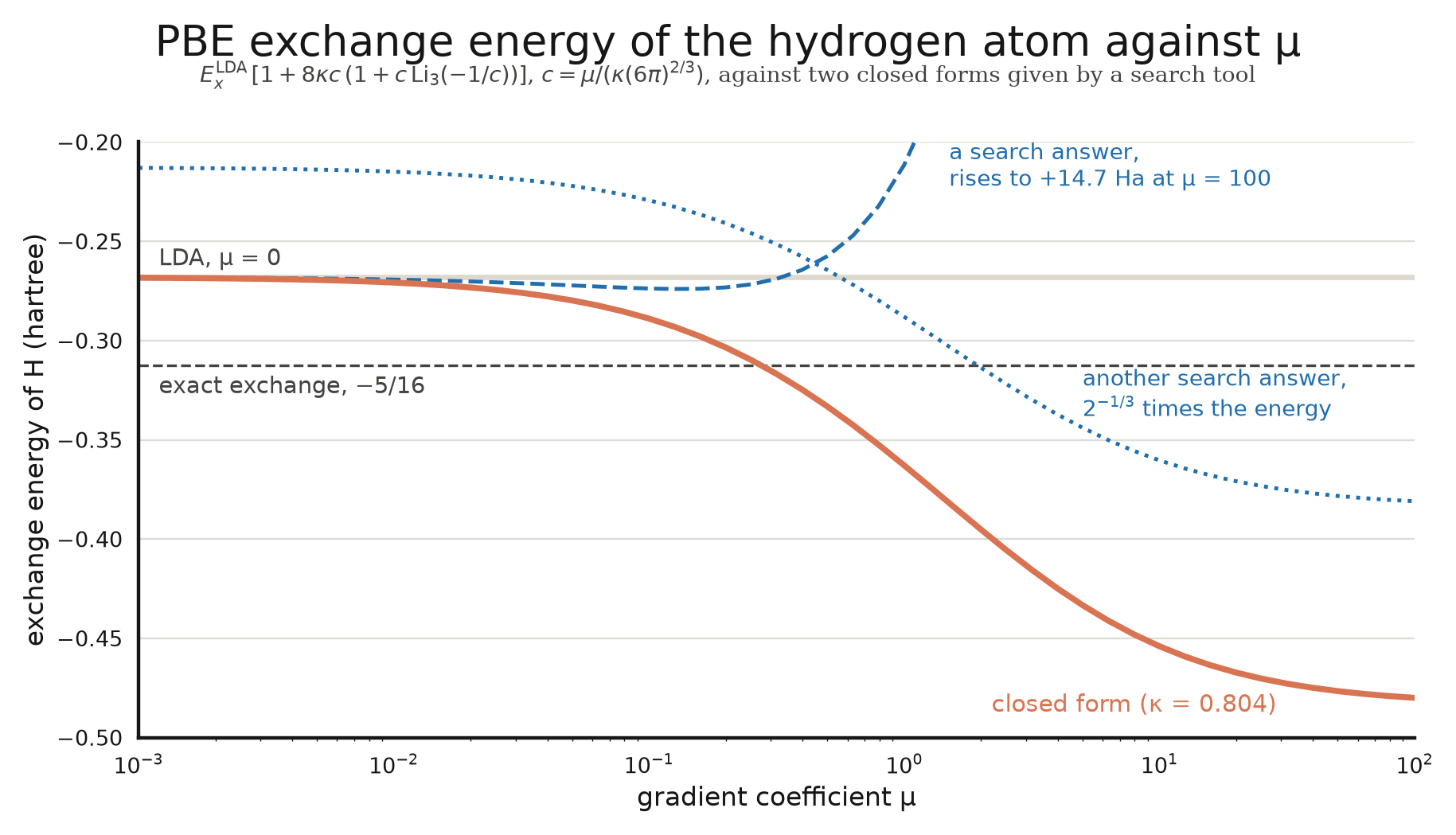}
\caption{The PBE exchange energy of hydrogen against $\mu$ at $\kappa = 0.804$, with two closed forms given by a search assistant (Section~\ref{sec:search}).}
\end{figure}

\paragraph{Checks and literature.} At 30 digits the closed form equals the radial integral of the definition (gap $0$) at three $(\kappa,\mu)$, and the two forms of $G$ agree with the integral to $1.1\times10^{-27}$ for $0.0385 \le c \le 20$. The value $-0.306$~hartree is published from numerical integration~\cite{he,peverati}; we found no closed form.

\FloatBarrier
\subsection{Short-range LDA exchange of hydrogen and of $n_0e^{-ar^p}$}

Range-separated hybrids such as $\omega$B97X-V and $\omega$B97M-V use the erf-attenuated LDA exchange of Gill, Adamson and Pople~\cite{gap}, whose attenuation in $q = k_F^2/\omega^2$ is $F_t(q) = \frac q9\, {}_2F_2(1,\frac32;\frac52,4;-q)$.

\paragraph{Result.} For hydrogen, with $z = (6\pi)^{2/3}/\omega^2$, \el{} returned a sixth-order operator and
\begin{equation}
\frac{E_x^{\rm sr}(\omega)}{E_x^{\rm LDA}} = \frac{8z}{243}\; {}_5F_5\!\left(1, \tfrac32, 3, 3, 3;\ \tfrac52, 4, 4, 4, 4;\ -z\right),
\end{equation}
which is $-0.14501734880753675451$~hartree at $\omega = 0.3$. In the variable $\omega$ itself two runs found nothing; the change to $z$ was ours. For any one-electron density $n_0e^{-ar^p}$ the ratio depends only on $j = 3/p$ and $w = (6\pi^2n_0)^{2/3}/\omega^2$. Given tables at $j = 1, 2, 4, 5$, the algorithm returned $ {}_3F_3$, $ {}_4F_4$, $ {}_6F_6$ and $ {}_7F_7$ forms, which with hydrogen give the pattern
\begin{equation}
R_j(w) = \frac{2^j w}{3^{j+2}}\; {}_{j+2}F_{j+2}\!\left(1, \tfrac32, 3, \ldots, 3;\ \tfrac52, 4, \ldots, 4;\ -w\right),\qquad R_j = \left(1 + \tfrac{\theta}{2}\right)R_{j+1},\quad \theta = w\frac{d}{dw},
\end{equation}
with $j$ threes and $j+1$ fours, annihilated by $(\theta-1)(2\theta+1)(\theta+2)^{j+1} + w\theta(2\theta+1)(\theta+2)^{j}$. The Mellin transform of $1 - R_j$ is $2^{j+1}\Gamma(s)/((1-2s)(1-s)(2-s)^{j+1})$, which gives the expansion at small $\omega$; for hydrogen its $w^{-2}$ term contains $\zeta(3)$. For the Gaussian, $j = 3/2$, the series coefficients carry $(m+3)^{-3/2}$, so by Flajolet, Gerhold and Salvy~\cite{fgs} the ratio satisfies no linear differential equation with polynomial coefficients; the algorithm found none for it at about 300 digits, nor for Hartree--Fock densities of N$_2$, O$_2$ and CO$_2$ in a Gaussian basis.

\begin{figure}[!htb]
\centering
\includegraphics[width=0.75\linewidth]{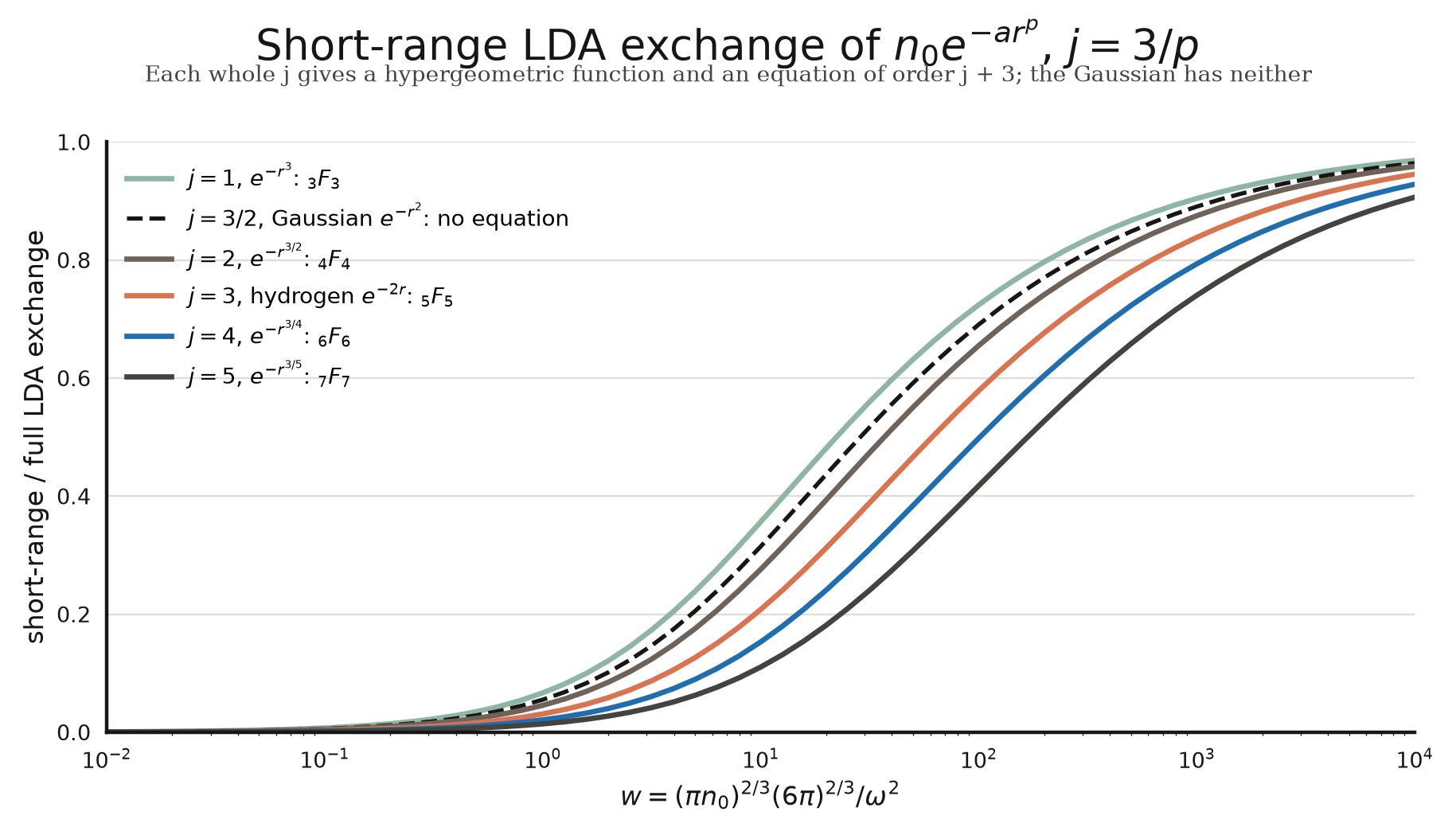}
\caption{The ratio $R_j(w)$ for $j = 1$ to $5$ and for the Gaussian, which has no equation.}
\end{figure}

\paragraph{Checks and literature.} At 30 digits the closed forms match the radial integral of the definition to $4\times10^{-31}$ at three $w$ for $j = 1$ to $5$, and the 130-digit tables to $2.3\times10^{-131}$. Gill, Adamson and Pople~\cite{gap}, Iikura et al.~\cite{iikura}, Toulouse, Colonna and Savin~\cite{tcs} and Toulouse's review~\cite{toulouse} define and study the functional without evaluating it in closed form for any density; we screened the 270 works citing~\cite{gap} and the 519 citing~\cite{tcs} by title.

\FloatBarrier
\subsection{Quasi-periods for the six-Bessel determinant at level 24}

Broadhurst and Roberts~\cite{br} conjectured that the moments $M(a,b,c) = \int_0^\infty I_0^a(t)K_0^b(t)\,t^c\,dt$ satisfy
\begin{equation}
\det\begin{pmatrix} M(0,6,0) & 3M(0,6,2) - 8M(0,6,4)\\ M(1,5,0) & 3M(1,5,2) - 8M(1,5,4)\end{pmatrix} = \frac{5\pi^6}{16},
\end{equation}
related the first column to the periods $P_m = \int_0^\infty g(iy)y^m\,dy$ of the weight-6 newform $g = q - 9q^3 - 34q^5 - \dots$ of level 24 (LMFDB 24.6.a.a), and wrote that they had not related the second column to Eichler integrals of a weakly holomorphic modular form.

\paragraph{Result (Exact Learning + PSLQ).} Let $h = -q + 8q^2 - 15q^3 + \dots$ be the cusp form of weight 6 on $\Gamma_0(24)$ fixed by the Atkin--Lehner involutions $W_3$, $W_8$, $W_{24}$, let $U = (\eta_1\eta_3\eta_8\eta_{24}/(\eta_2\eta_4\eta_6\eta_{12}))^2$ with $\eta_k = \eta(kz)$, and let
\begin{equation}
G = (T_5 - 54)(T_5 + 66)(T_5 - 3126)\,[h\,U^2],
\end{equation}
the part of $hU^2$ with $T_5$ eigenvalue $-34$. $G$ is holomorphic on the upper half plane, vanishes at the cusps $0$ and $\infty$ and has poles at $1/2, 1/4, 1/6, 1/12$, so its Eichler integrals $A_m = \int_0^\infty G(iy)y^m\,dy$ converge. With $B_m = P_m + A_m/1334784$,
\begin{equation}
3M(1,5,2) - 8M(1,5,4) = \frac{\pi^5}{432}B_0,\quad 3M(0,6,2) - 8M(0,6,4) = \frac{\pi^6}{12}B_1,\quad P_0B_1 - B_0P_1 = \frac{45}{4\pi^5}.
\end{equation}
The last is a Legendre relation between the periods and the quasi-periods of $g$, and with the first column it gives the determinant.

\begin{figure}[!htb]
\centering
\includegraphics[width=0.85\linewidth]{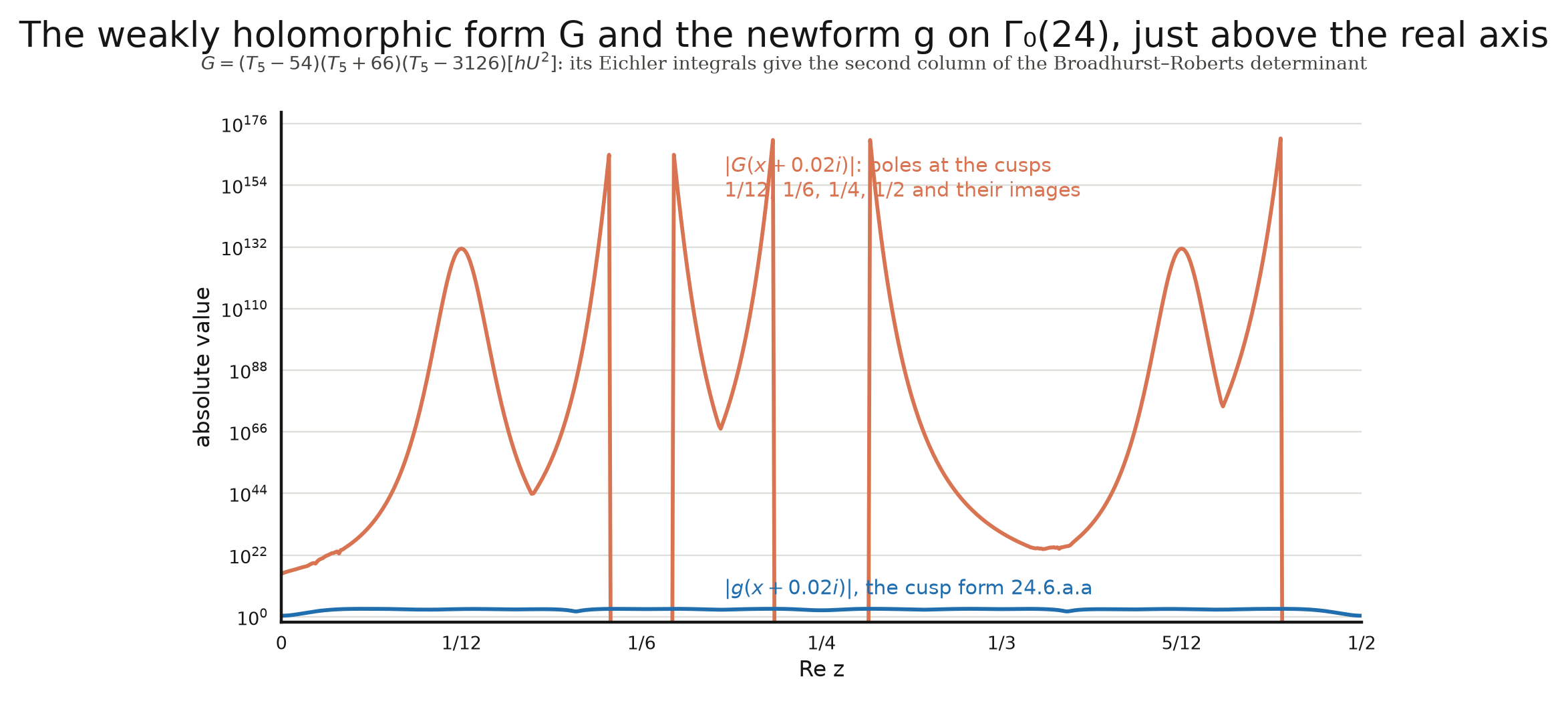}
\caption{$|G|$ and $|g|$ on the line $\operatorname{Im} z = 0.02$: $G$ has poles at the cusps $1/12$, $1/6$, $1/4$, $1/2$ and their images, while the cusp form $g$ stays small.}
\end{figure}

\paragraph{How it was found and checked.} Integer relation searches on the moments at 90 digits gave the linear relations $M(0,6,2) - 3\pi^2M(2,4,2) = -\pi^4/16$ and $M(0,6,4) - 3\pi^2M(2,4,4) = -3\pi^4/128$ and the Legendre relation in the moments. The search for the quasi-periods ran over the forms $hU^2$ and $hU^3$ for each cusp form $h$ fixed by $W_3$ and $W_8$, projected onto $T_5 = -34$, with PSLQ of the second column against $\pi^sA_m$ and $\pi^sP_m$. An independent recomputation at 40 digits (moments by tanh--sinh quadrature; forms, periods and L-values with PARI/GP~\cite{pari}) confirms every relation to a relative gap below $10^{-35}$.

\paragraph{Literature.} Acres and Broadhurst~\cite{ab} give quasi-periods from weakly holomorphic eta quotients at level 6 for moments with odd powers of $t$; Zhou~\cite{zhou} and Fres\'an, Sabbah and Yu~\cite{fsy} prove the quadratic relations for odd moments, and Chuang and Yu~\cite{cy} prove linear and quadratic relations for the even moments, the setting of this determinant, without modular forms. The determinant very likely follows from~\cite{cy}; what we add is the weakly holomorphic form. We screened the 22 works citing~\cite{br}. The four-Bessel analogue that the same search found, $\int_0^\infty t^2K_0^4\,dt$ as a period of a weight-4 eta quotient on $\Gamma_0(8)$, is equivalent to~\cite[(4.13)]{br}.

\FloatBarrier
\subsection{Exact states of a harmonic trap with soft Coulomb terms}

In $d$ dimensions, with $z = \sqrt{1 + \rho^2}$, the Schr\"odinger equation $-\frac12\nabla^2\psi + V\psi = E\psi$ for
\begin{equation}
V = \tfrac12 w^2\rho^2 + w\kappa z + \frac{\lambda}{z} + \frac{\mu}{z^2}
\end{equation}
has exact states $\psi = \rho^l h(z)e^{-wz^2/2-\kappa z}$, $E = w(n + D/2) - \kappa^2/2$, $D = d + 2l$, with $h$ a polynomial of degree $n$.

\paragraph{Result.} For $\mu = 0$ and $n = 1$ there is a family in every dimension,
\begin{equation}
h = z + \frac1\kappa,\qquad w = \frac{\kappa^2(\kappa^2 - D - 1)}{2(\kappa^2 - 1)},\qquad \lambda = -\frac{\kappa^3}{2},\qquad E = -\frac{D\kappa^2(D + 3 - \kappa^2)}{4(\kappa^2 - 1)},
\end{equation}
and for $\mu \ne 0$, which forces $h = z^2q(z)$, a second family at $n = 3$. At $n = 2$ and $n = 4$ there are rational points. With $\kappa < 0$ the potentials are rings in two dimensions and double wells in one: $\kappa = -2$ gives the ring $V = \frac29\rho^2 - \frac43z + \frac4z$ with ground state $(2z-1)e^{-z^2/3+2z}$ at $E = -\frac23$. For $\kappa = \mu = 0$, two conditions fix $(w,\lambda)$ at isolated algebraic points, which are exact states of two electrons in a harmonic trap interacting by $1/\sqrt{a^2 + u^2}$, analogues of Taut's states~\cite{taut} for the bare Coulomb interaction.

\begin{figure}[!htb]
\centering
\includegraphics[width=0.7\linewidth]{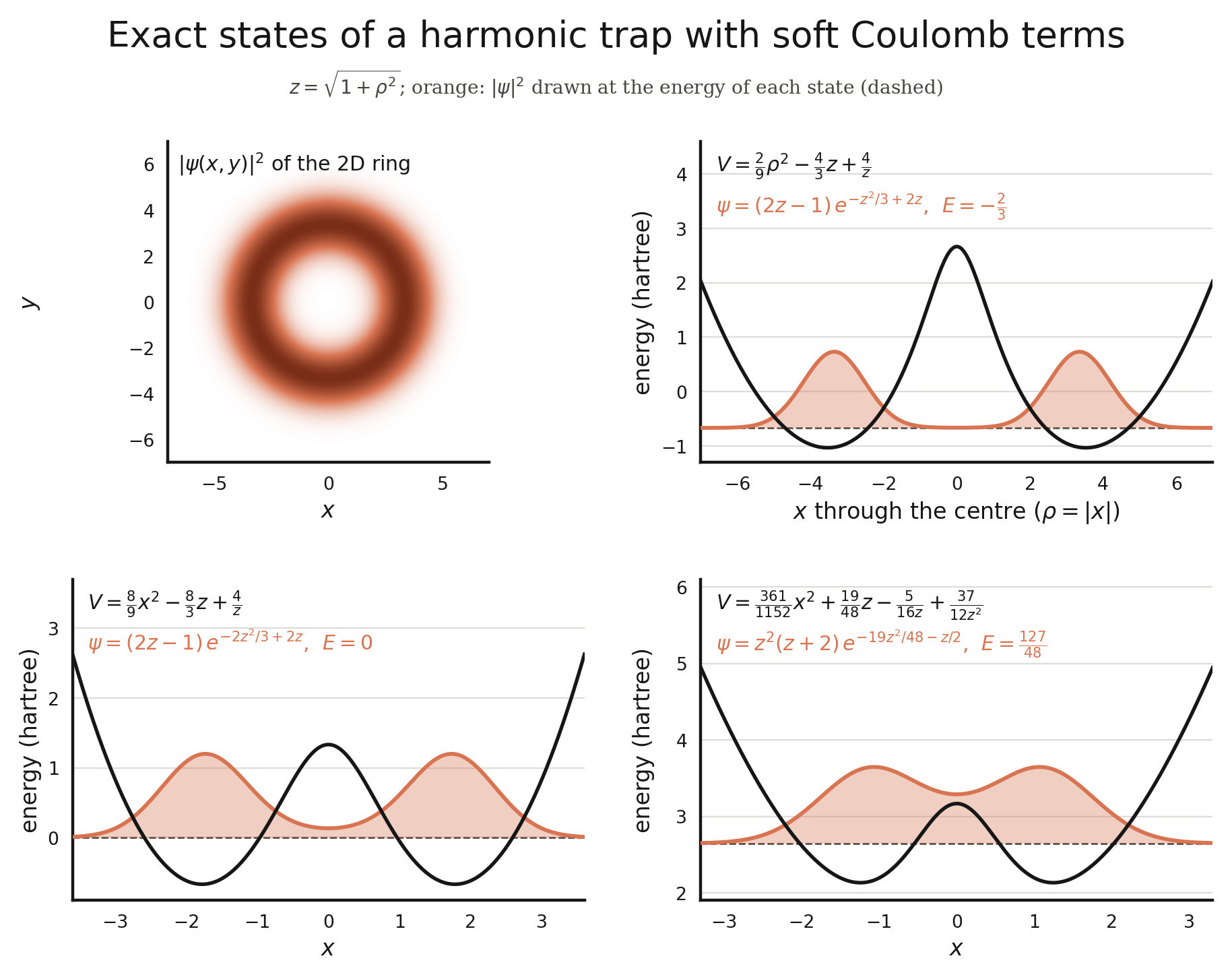}
\caption{The two-dimensional ring and the one-dimensional double well at $\kappa = -2$, and a one-dimensional state of the Lorentzian family at $\kappa = \frac12$. Orange: $|\psi|^2$ drawn at the energy of the state (dashed).}
\end{figure}

\paragraph{How it was found and checked.} \el{} was given 50-digit tables of $\psi$ from a radial solver run at 170 digits, and returned closed forms for 13 states, such as $(2z-1)e^{-z^2/3+2z-5/3}$ for the ring. These suggested the polynomial ansatz, whose conditions were solved with sympy, and the algorithm then recovered members of each family it had not been given. With sympy, $(H - E)\psi/\psi = 0$ holds identically for both families with $D$ and $\kappa$ symbolic and for every rational point; at the trap points the residual is below $10^{-59}$ at 60 digits.

\paragraph{Literature.} At $w = 0$ the problem is the soft Coulomb atom, solved in one~\cite{lc}, three~\cite{clark} and any dimension~\cite{hss}; at $\kappa = 0$ the Lorentzian problem is the non-polynomial oscillator~\cite{cm}. Hall and Saad~\cite{hs} treat a trapped problem with the softening $1/(r+\beta)$, and Agboola and Zhang~\cite{az1} and Hall, Saad and Bryenton~\cite{hsb} the same softening without a trap. Ko\'scik and Sowi\'nski~\cite{ks} solve two particles in a trap with a step interaction. We screened 309 citing works in a first search and the 25 works citing~\cite{hss}, and did not find states with $w > 0$ and the softening $\sqrt{1 + \rho^2}$.

\FloatBarrier
\subsection{The period of the planar $L_p$ dual Minkowski problem}

Positive solutions of $u^{1-p}(u'^2 + u^2)^{(q-2)/2}(u'' + u) = 1$ on $S^1$ are support functions of convex bodies. Li and Wan~\cite{lw} showed that closed solutions are counted by the crossings of a half-period $\Theta(p,q,r)$ with the levels $\pi/k$, and Tao~\cite{tao} completed the count by writing $\Theta$ as a convolution in a parameter $E$ and computing its Laplace transform, a ratio of Gamma functions times $C^s$ for a constant $C$ he did not name.

\paragraph{Result.} With $m = -p > 0$ and $\frac1\omega = \frac1m + \frac1q$, the constant is $C = 1/\beta$, $\beta = m^{1/q}q^{1/m}(m+q)^{-1/\omega}$, and
\begin{equation}
\Theta = \frac{\sqrt\pi}{2(q+m)}\;H^{3,0}_{2,3}\!\left[\beta e^{-E}\,\middle|\,\begin{matrix}(1, \frac12),\ (0, \frac1\omega)\\ (\frac12, \frac12),\ (0, \frac1m),\ (0, \frac1q)\end{matrix}\right],
\end{equation}
a Fox $H$ function whose residue series converges for every $E > 0$. For rational $p$ and $q$ Gauss's multiplication formula reduces it to $\Theta = \frac{\pi}{\sqrt{q-p}}G^{M,0}_{M,M}(e^{-LE}\,|\,a; b)$; for example
\begin{equation}
\Theta(-1, 2) = \frac{\pi}{\sqrt3}\;G^{3,0}_{3,3}\!\left(e^{-2E}\,\middle|\,\begin{matrix}\frac13, \frac23, 1\\ 0, \frac12, \frac12\end{matrix}\right).
\end{equation}
Li and Wan's dualities carry the formula to $p$ and $q$ of the same sign; at $p = 0$ the transform is not a ratio of Gamma functions and the algorithm found no equation.

\begin{figure}[!htb]
\centering
\includegraphics[width=0.8\linewidth]{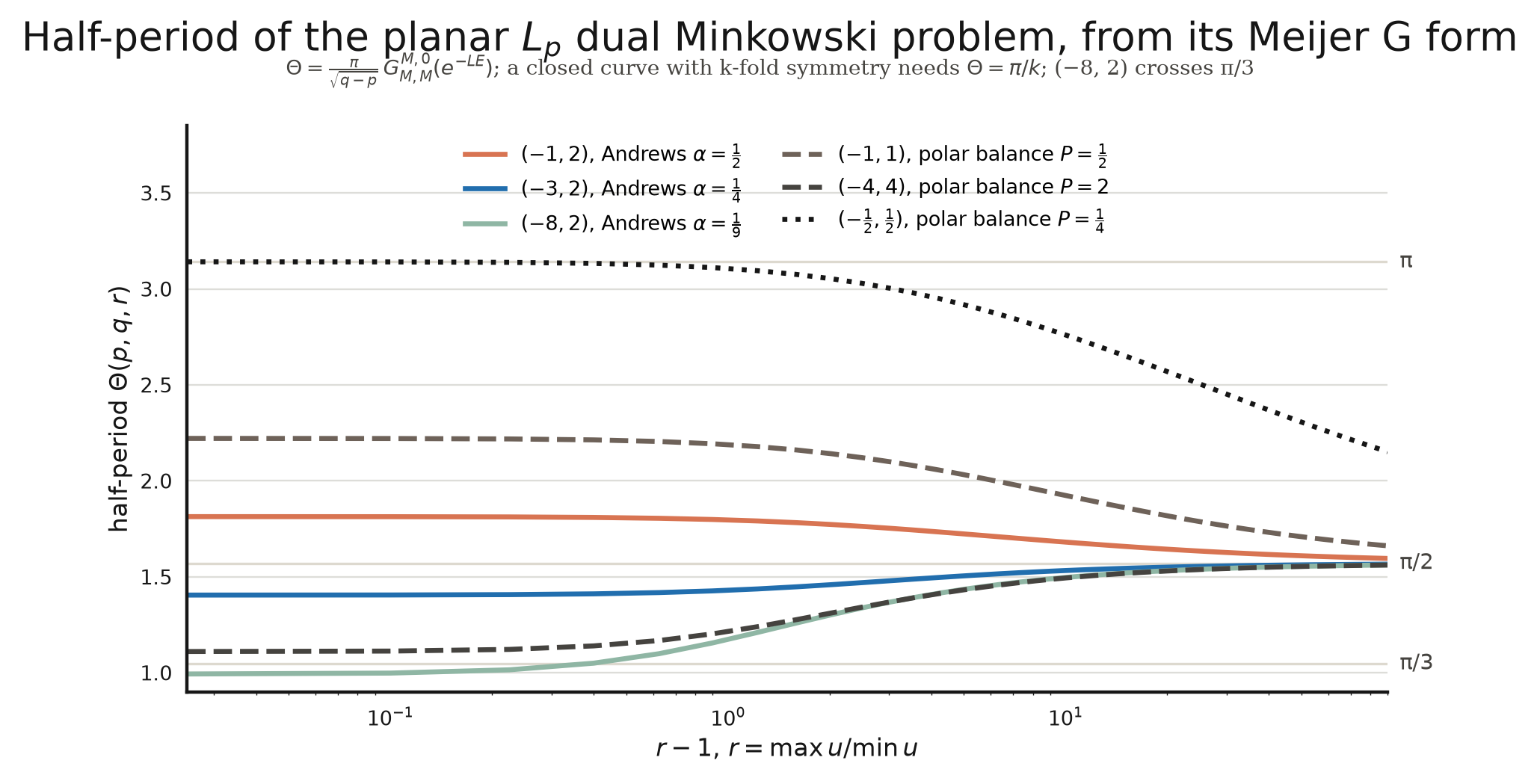}
\caption{The half-period against $r$ for six $(p,q)$ from the Meijer $G$ form, with the levels $\pi/k$.}
\end{figure}

\paragraph{How it was found and checked.} The work began on the line $q = -p$, where the equation describes bodies in balance with their polar. For nine rational $P = q/2$, \el{} returned an equation of order 3, 5 or 7 and a Meijer $G$ form for the period from tables of its integral; reading off their parameters gave the general form on the line, and a search then found Tao's transform, whose constant follows by matching. Off the line, at $(p,q) = (-1,2)$ and $(-3,2)$, the algorithm returned the same functions from 120- and 220-digit tables. At 30 digits the Meijer $G$ form matches Li and Wan's integral at thirteen rational $(p,q)$ and three $r$ each to $8.5\times10^{-31}$, and the Fox $H$ series at three irrational pairs to $4\times10^{-30}$.

\paragraph{Literature.} The inversion of Tao's transform is a standard step once the transform is known; what is recorded here is that step carried out, with the constant and the rational reduction, and the route by which the algorithm reached the same functions without it. We read~\cite{lw,tao,ll,andrews} and screened the 20 works citing~\cite{lw}, the 8 citing~\cite{ll} and the 191 citing~\cite{andrews}.

\FloatBarrier
\section{Known results recovered from numbers, and errors found}
\label{sec:known}

\FloatBarrier
\subsection{Feynman integrals}

Nineteen integrals of quantum field theory were computed to between 40 and 200 digits and given to \el{} without any information about the diagrams; it returned the equation of seventeen. In two dimensions the banana integral with $n$ massive lines, $f_n(x) = \int_0^\infty tJ_0(\sqrt xt)K_0(t)^n\,dt$, came back at order 3 for $n = 3$ from its formula and at orders 4 to 6 for $n = 4$ to $6$ from tables of 641 values at 105 to 200 digits. Each operator is $D\mathcal{L}_n$, the row $\mathcal{L}_n$ of Vanhove's Table~1~\cite{vanhove} differentiated once; the three-loop row is the theorem of Bloch, Kerr and Vanhove~\cite{bkv}. The vacuum integrals $V_n(x) = \int_0^\infty tK_0(t)^nK_0(xt)\,dt$ follow $\theta^2\mathcal{L}_n$ with $\theta = x\,d/dx$, and $V_2$ has a Clausen form equal to that of Davydychev and Tausk~\cite{dt}. The two-loop contribution of a lighter lepton loop to a magnetic moment came back as a Meijer $G^{4,2}_{4,4}$ equal to Elend's formula as printed by Passera~\cite{passera} to $2\times10^{-40}$.

\begin{figure}[!htb]
\centering
\includegraphics[width=\linewidth]{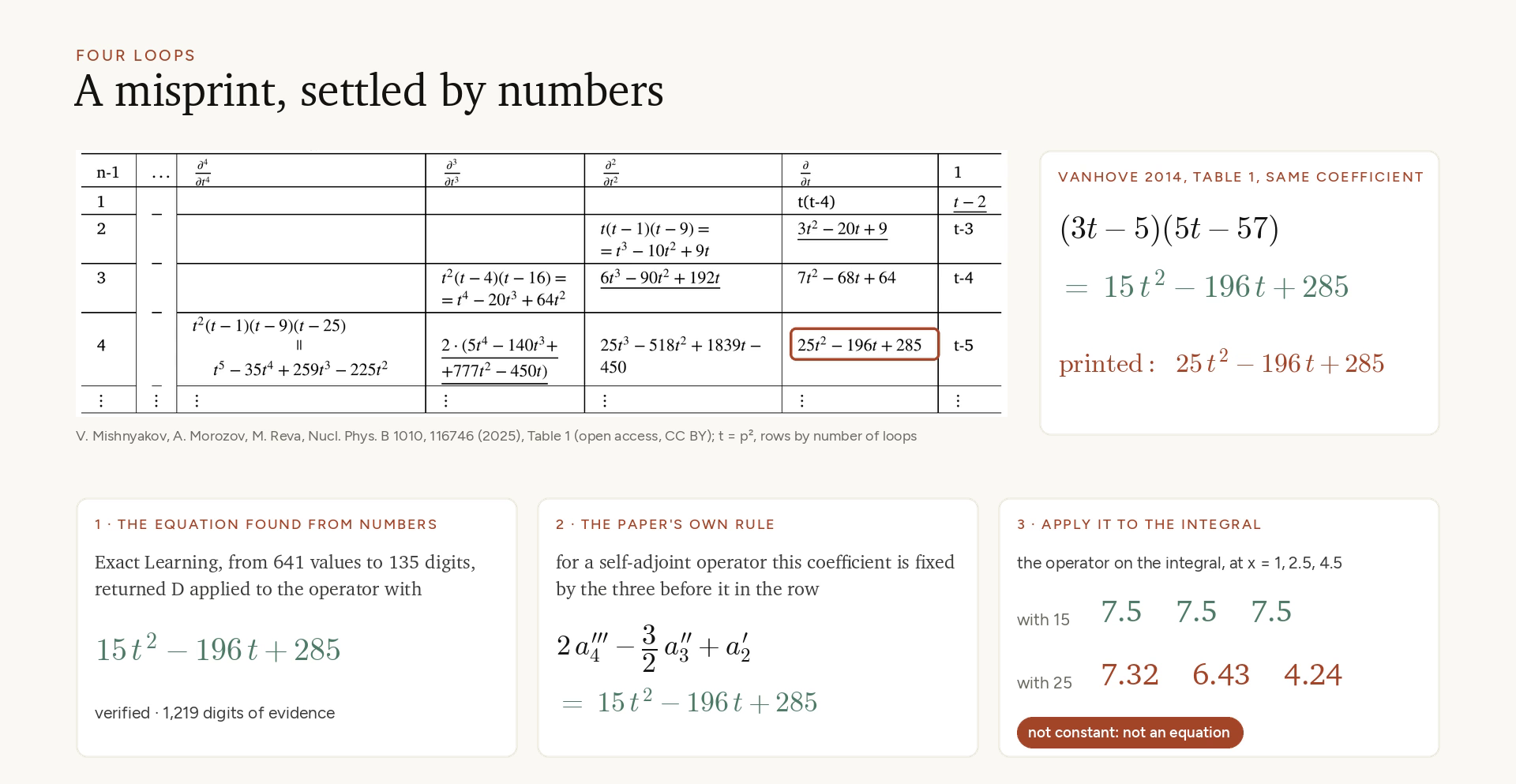}
\caption{The four-loop row of~\cite[Table 1]{mmr} (reproduced under CC BY) and three checks that give $15t^2$ for the printed $25t^2$.}
\label{fig:misprint}
\end{figure}

\paragraph{A misprint.} Mishnyakov, Morozov and Reva~\cite{mmr} reprint these equations in their Table~1. Their four-loop row prints the coefficient of $\partial_t$ as $25t^2 - 196t + 285$; Vanhove has $(3t-5)(5t-57) = 15t^2 - 196t + 285$. The operator returned from the 135-digit table has 15; the paper's own condition for a self-adjoint operator, $2a_4''' - \frac32a_3'' + a_2'$, gives 15; and the row applied to the integral is the constant $7.5$ at $x = 1, 2.5, 4.5$ with 15 and $7.32, 6.43, 4.24$ with 25. The misprint is in arXiv:2311.13524v1 and in the journal version (Figure~\ref{fig:misprint}).

\paragraph{No model, and why.} The sunrise with masses $1, 1, 3$ returned no model. Its second-order equation from M\"uller-Stach, Weinzierl and Zayadeh~\cite{mwz} has $\log 3$ on the right-hand side, so the third-order operator that removes it has a coefficient outside the rationals; split into two integrals without the logarithm, each came back at third order. Version 1 of~\cite{mwz} has two wrong exponents in the coefficient of the logarithm, corrected in version 2.

\FloatBarrier
\subsection{Lattice Green functions}

\begin{figure}[!htb]
\centering
\includegraphics[width=\linewidth]{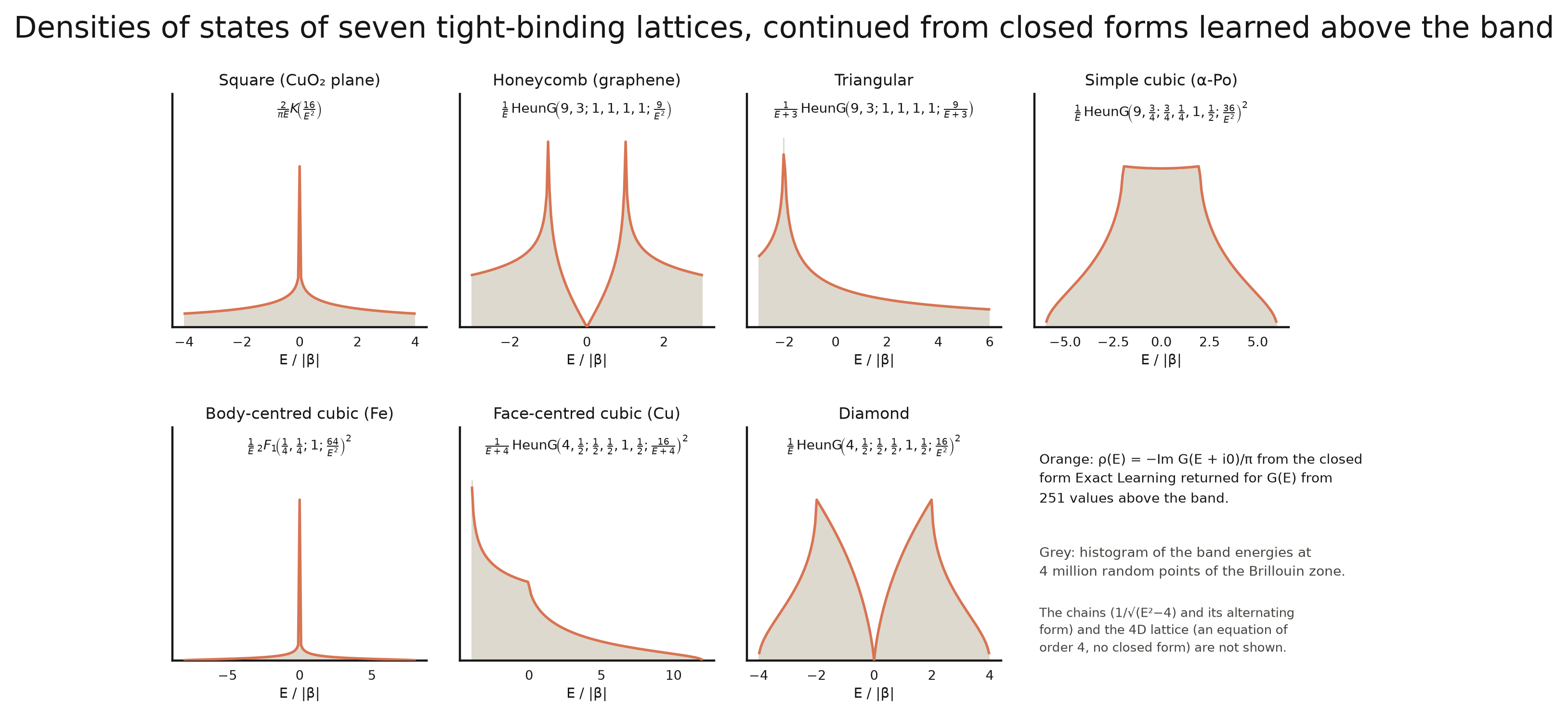}
\caption{Densities of states continued from the closed forms (orange) against histograms of the band energies (grey).}
\label{fig:lattice}
\end{figure}

The lattice Green function $G(E) = \langle 0|(E - H)^{-1}|0\rangle$ of ten tight-binding lattices, from the chain to the four-dimensional hypercubic lattice, was given to \el{} as 251 values above the band, summed from closed-walk counts. One call per lattice returned the equation of all ten and the closed form of the nine below four dimensions, for example
\begin{equation}
G_{\rm fcc}(E) = \frac{1}{E+4}\,\mathrm{HeunG}\!\left(4, \tfrac12; \tfrac12, \tfrac12, 1, \tfrac12; \frac{16}{E+4}\right)^{2},
\end{equation}
and for the four-dimensional lattice the operator of Glasser and Guttmann~\cite{gg}. Every result is published, from Watson~\cite{watson} to Guttmann's review~\cite{guttmann}. Continued into the band, the closed forms give densities of states that agree with histograms of the band energies at four million random points of the Brillouin zone (Figure~\ref{fig:lattice}).

\paragraph{Errata.} Evaluated as printed against the walk counts at 90 digits, four of the three-dimensional forms in~\cite{guttmann} (arXiv v1, eqs.~(20)--(23)) each carry one wrong parameter: for the simple cubic lattice the accessory parameter $-\frac34$ should be $+\frac34$, as in his Heun equation (19); for bcc $ {}_2F_1(\frac14, -\frac14; 1; z^2)$ should be $ {}_2F_1(\frac14, \frac14; 1; z^2)$; for fcc $\delta = 1$ should be $\frac12$; and for diamond the accessory parameter $-\frac12$ should be $+\frac12$. As printed they are 32\%, 25\%, 3.7\% and 38\% off one unit above the band; corrected they agree to $10^{-81}$ or better and match the forms returned by the algorithm.

\FloatBarrier
\subsection{Other corrections}
\label{sec:pf}

\paragraph{The eccentricity expansion of the ellipse density.} Parry and Fischbach~\cite{pf} expand $f(r,b)$ in $\epsilon = \sqrt{1-b^2}$ through $\epsilon^4$, split at $r = 2b$. Their inner series (A2) is exactly the expansion of the Kamp\'e de F\'eriet form, with error $O(\epsilon^6)$. Their outer series (A3) does not converge to $f$: the outer piece is a band of width about $2\epsilon^2$ next to the branch point of $f_{\rm disk}$, every term of an expansion at their fixed $(r, X)$ is of the order of $f$ itself, and from the $F_3$ form $f = \frac{2\epsilon^3}{\pi}(1-X^2)^2\, {}_2F_1(\frac12,\frac12;3;1-X^2)(1 + O(\epsilon^2))$, from which (A3) stays 3\% to 5\% away as $\epsilon \to 0$. Expanding~\eqref{eq:ell-avg} at fixed angle to the same order reproduces the leading behaviour of (A3) to $5\times10^{-5}$, so the error lies in the expansion, not in a misprint. A web page on the moments of the ellipse~\cite{behrend} prints $a^3$ for $a^5$ in the fifth moment.

\paragraph{A missing term.} Agboola and Zhang~\cite{az2} print their $q = 2$ Klein--Gordon equation for $\chi$ (eq.~(2.29) of arXiv:1311.1566v1) with $-\xi\beta^2\chi$, where the reduction of their radial equation (2.23) gives $-(\xi\beta^2 + \xi^2\beta^2t)\chi$. Their $n = 1$ state (4.41) then leaves a residual of order one in (2.23) at 40 digits; with the term restored it is the soft Coulomb state of Clark~\cite{clark}. We have not seen the journal version.

\FloatBarrier
\section{Searches and the answers of a search assistant}
\label{sec:search}

\begin{table}[ht]
\centering
\small
\begin{tabular}{>{\raggedright}p{4.0cm}>{\raggedright}p{2.6cm}c>{\raggedright\arraybackslash}p{5.6cm}}
\toprule
Result & Method & Digits & Literature read or screened \\
\midrule
Ellipse distance: ODE, PDE, $F_3$, KdF & Exact Learning & 30--60 & Kellerer, Parry--Fischbach and their 77 + 28 citing works; Lellouche--Souris, Garc\'ia-Pelayo, Sheng, Ge\v{c}iauskas \\
PBE exchange of H & Exact Learning & 30 & PBE; He; Peverati \\
Short-range LDA exchange, $j = 1$--5 & Exact Learning & 30--131 & Gill--Adamson--Pople (270 citing), Toulouse--Colonna--Savin (519 citing, by title) \\
Six-Bessel quasi-periods & Exact Learning + PSLQ & 35 & Broadhurst--Roberts (22 citing), Acres--Broadhurst, Zhou, Fres\'an--Sabbah--Yu, Chuang--Yu \\
Soft Coulomb trap & Exact Learning & 50--60 & 309 works in a first search; Hall--Saad--Sen (25 citing); Agboola--Zhang; Ko\'scik--Sowi\'nski \\
$L_p$ dual Minkowski period & Exact Learning & 30 & Li--Wan (20 citing), Liu--Lu (8), Andrews (191), Tao \\
\midrule
Feynman integrals & Exact Learning & 20--40 & known: Vanhove; Bloch--Kerr--Vanhove; M\"uller-Stach--Weinzierl--Zayadeh \\
Lattice Green functions & Exact Learning & 81--90 & known: Watson; Joyce; Glasser--Guttmann; Guttmann \\
\bottomrule
\end{tabular}
\caption{The results, how they were found, the precision of the numerical checks, and the literature read or screened. The six above the rule were not found in the sources searched.}
\label{tab:summary}
\end{table}

\begin{table}[ht]
\centering
\small
\begin{tabular}{>{\raggedright}p{4.4cm}>{\raggedright}p{4.6cm}>{\raggedright\arraybackslash}p{4.6cm}}
\toprule
Source & As printed & Corrected \\
\midrule
Mishnyakov--Morozov--Reva~\cite{mmr}, Table 1, four loops & $25t^2 - 196t + 285$ & $15t^2 - 196t + 285$ \\
Guttmann~\cite{guttmann}, (20) simple cubic & accessory parameter $-\frac34$ & $+\frac34$ \\
Guttmann~\cite{guttmann}, (21) bcc & $ {}_2F_1(\frac14, -\frac14; 1; z^2)^2$ & $ {}_2F_1(\frac14, \frac14; 1; z^2)^2$ \\
Guttmann~\cite{guttmann}, (22) fcc & $\delta = 1$ & $\delta = \frac12$ \\
Guttmann~\cite{guttmann}, (23) diamond & accessory parameter $-\frac12$ & $+\frac12$ \\
Parry--Fischbach~\cite{pf}, (A3) & series in $\epsilon$ for $2b < r < 2$ & does not converge to $f$; use the $F_3$ form \\
Agboola--Zhang~\cite{az2}, (2.29) & $-\xi\beta^2\chi$ & $-(\xi\beta^2 + \xi^2\beta^2t)\chi$ \\
M\"uller-Stach et al.~\cite{mwz} v1 & $-5m_1^5m_3^2$, $+3m_1^2m_3^4$ & $-5m_1^6m_3^2$, $+3m_1^4m_3^4$ (as in v2) \\
Behrend~\cite{behrend}, fifth moment & $a^3$ & $a^5$ \\
\bottomrule
\end{tabular}
\caption{Errors in published sources found by comparison with the numbers.}
\label{tab:errata}
\end{table}

\paragraph{Answers of a search assistant.} When asked for these closed forms, a web search assistant mostly said that none exists, or gave a formula that the definition does not support. For PBE exchange of hydrogen its answers included an integral missing the $r^2$ of the volume element, a coefficient $-2c$ for $-2c^2$ with the unpolarised density, a prefactor $2^{-1/3}$ too small, the RPBE enhancement factor in place of PBE's, and the values $-0.316$ and $-0.384$~hartree, both below the exact $-0.3125$. For short-range exchange it gave $-0.194$~hartree at $\omega = 0.3$, where the value is $-0.145$, and said the result is not a single hypergeometric function. For the ellipse it gave the density of the disk multiplied by four, and a second-order equation that the disk density does not satisfy. For the six-Bessel moments it swapped the normalisations of $L(g,4)$ and $L(g,5)$, and gave $\int_0^\infty tI_0K_0^5\,dt = 144L(g,5)$, which holds without the factor $t$. We list these because they are what a search for the results currently returns.

\FloatBarrier
\section{Conclusion}

Each result in Section~\ref{sec:new} began as a table of numbers, and in each the algorithm returned an exact equation from which the closed form followed, by its own solver or by a short reduction. The same procedure, applied to known results, settled misprinted coefficients by evaluation alone. Later issues will report further results in the same format.

\end{document}